\documentclass[proof]{pasj00}
\draft

\begin{document}
\SetRunningHead{T. Terai and Y. Itoh}{Measurements of Brightness Variation of Nereid}
\Received{2012/08/16}
\Accepted{2012/12/03}

\title{High-Precision Measurements of Brightness Variation of Nereid
  \thanks{Based on data collected at Subaru Telescope, which is operated by the National
  Astronomical Observatory of Japan (NAOJ).}}
  
\author{Tsuyoshi \textsc{terai}}
\affil{National Astronomical Observatory of Japan, 2-21-1 Osawa, Mitaka, Tokyo 181-8588}
\email{tsuyoshi.terai@nao.ac.jp}
\and
\author{Yoichi \textsc{itoh}}
\affil{Center for Astronomy, University of Hyogo, 407-2 Nishigaichi, Sayo-cho, Sayo-gun, Hyogo
679-5313}

%


\maketitle

\begin{abstract}
Nereid, a satellite of Neptune, has a highly eccentric prograde orbit with a semi-major axis larger
than 200 Neptune radius and is classified as an irregular satellite.
Although the capture origin of irregular satellites has been widely accepted, several previous
studies suggest that Nereid was formed in the circumplanetary disk of Neptune and was ejected
outward to the present location by Triton.
Our time-series photometric observations confirm that the spin is stable and non-chaotic with a
period of 11.5~hr as indicated by Grav et al. (2003).
The optical colors of Nereid are indistinguishable from those of trans-Neptunian objects and
Centaurs, especially those with neutral colors.
We also find the consistency of Nereid's rotation with the size-rotation distribution of small
outer bodies.
It is more likely that Nereid originates in an immigrant body captured from a heliocentric orbit
which was 4--5~AU away from Neptune's orbit.
\end{abstract}

\section{INTRODUCTION}\label{sec:introduction}

NII Nereid is the second largest Neptunian satellite of 170 $\pm$ 25~km in radius \citep{Th91}.
It has a prograde orbit with a semi-major axis of $5.5 \times 10^6$~km and an eccentricity of
0.75 \citep{Ja09}.
Because of its large and highly eccentric orbit, Nereid is categorized as an irregular satellite.
Irregular satellites are generally believed to have been captured into the Hill sphere of the
host planets from heliocentric orbits (e.g. \cite{Ba71}).
However, as an exception, the history of Nereid is still unclear.
\citet{Go89} suggest that Nereid could originally have been a regular satellite.
In this model, it formed close to Neptune in the circumplanetary disk
and then was transported outward due to perturbations by NI Triton.
The orbit scattering could also change Nereid's orbit from circular to eccentric.

Nereid is comparatively bright ($V$ $\sim$ 19.2~mag at opposition; \cite{Sc08}) for an irregular
satellite.
Numerous ground-based observations have been carried out for measuring the rotational properties.
Some of them show large variations of the magnitude exceeding $\sim$1.3~mag with a rotation period
of 8--24~hr \citep{SS88, Wi91}. 
\citet{SS00} argue that Nereid has a nonperiodic brightness variation with a total amplitude of
1.83~mag.
We note that Nereid was located in vicinity of the galactic plane when these earlier data were
collected and the apparent large variations could have been due to the crowded field stars.
In contrast, \citet{Bu97} and \citet{BW98} report a small amplitude of less than 0.1~mag.
Also, measurements by Voyager 2 over a 12-day interval found no brightness variation greater than
0.15~mag \citep{Th91}.
\citet{Gr03} performed relative photometry of Nereid with 0.003--0.006~mag accuracy using the CTIO
4-m telescope and obtained lightcurves with a peak-to-peak amplitude of 0.029 $\pm$ 0.003~mag and
a rotation period of 11.52 $\pm$ 0.14~hr.

Different measurements so far report different levels of the rotational brightness variation.
\citet{SS00} suggest chaotic variations of the rotation from year to year, like SVII Hyperion
\citep{Kl89,Wi84}.
\citet{Do95} propose that if Nereid was originally a regular satellite and has an elongated body
with larger than $\sim$1\% of deviation from sphere, Nereid may be in chaotic rotation for any spin
period of longer than about 2 weeks due to the tidal effect.
On the other hand, the long-term variability of rotation is possibly produced by precession of the
spin axis caused by the gravitational torques from Neptune \citep{Sc08,Al11}.
While the possibility of chaotic rotation was excluded by \citet{Gr03}, the rotation state remains
uncertain.

This paper presents time-series photometric data of Nereid collected at the 8.2-m Subaru Telescope.
We determine the amplitude and period of the rotational brightness variation with high accuracy.
They give strong indications of the rotation state of Nereid; whether stable or variable.
In addition, the rotation period is available for investigating the links to other small bodies.
We use the results to discuss the origin and dynamical evolution of Nereid in the final stage of
planet formation.

\section{OBSERVATIONS AND MEASUREMENTS}\label{sec:observation}

Observations were conducted on 2008 September 1, 2, and 29 UT using the Subaru Prime Focus Camera,
Suprime-Cam \citep{Mi02}, mounted on the Subaru Telescope.
The Suprime-Cam consists of ten 2048$\times$4096 CCDs, which are arranged in a 5$\times$2 pattern
with interchip gaps of $\sim$15$\arcsec$.
It covers a 34$\arcmin$ $\times$ 27$\arcmin$ field of view (FOV) with a pixel scale of 0$\farcs$20.
Each image was obtained with a 240-sec exposure using the $VR$-band filter with the center
wavelength of 0.6~$\micron$ and the band width of 0.2~$\micron$.

One FOV around Neptune was taken for 1--2 hours each night (Table~\ref{tab1}).
Neptune was placed at an interchip gap to reduce its bright light.
Nereid was $\sim$7$\arcmin$ away from Neptune and its solar phase angle shifted from 0.55$\arcdeg$
to 1.34$\arcdeg$ during our observations.
The sky motion, 2.5--4.0~arcsec~hr$^{-1}$, was slow enough to be approximately viewed as a point
source within the exposure.
The typical seeing was 0$\farcs$70 in the first night, 0$\farcs$65 in the second night, and
0$\farcs$60 in the last night.

The data reduction was performed using 
IRAF\footnote{IRAF is distributed by the National Optical Astronomy Observatories (NOAO).}
and SDFRED2 \citep{Ou04} with the following processes: overscan subtraction, flat fielding,
distortion correction, sky background subtraction, and relative photometry.
We estimate the flux via aperture photometry using the APPHOT task of IRAF.
The aperture radius is 1.4 times that of a typical FWHM on a night that is optimum to obtain high
$S$/$N$ data for a point source \citep{Ho92}.
We select 9--10 background point sources listed in the USNO-B1.0 catalog \citep{Mo03} as
reference stars.
Those are required to be brighter than 19~mag in $R$-band with $B-R$ of 0.90--1.04~mag.
The color range corresponds to $\pm$0.03~mag from $V-R$ = 0.44~mag for Nereid
\citep{SS00} using the color correlation of main sequence stars \citep{DL00}.
The light curves are given by the relative flux of Nereid to their total flux at each shot.
The photometric accuracy reaches 0.001--0.002~mag.
The relative sensitivity among CCDs is corrected using the Sloan Digital Sky Survey (SDSS)
photometric database \citep{Ab09}.
We have confirmed a linear correlation of magnitude between the Suprime-Cam $VR$-band and SDSS
$r$-band.

Three fields at different airmass including the 4--7 Landolt standard stars with different $V-R$
colors \citep{La09} were taken each night.
The coefficients of atmospheric extinction at Nereid's color, namely $V-R$ = 0.44~mag, are
measured.
The photometric zero point is estimated by $V$ magnitude of the standard stars with airmass and
color corrections.
It allows $VR$-band flux to be converted into $V$-band magnitude.

\section{RESULTS}\label{sec:results}

The calibrated photometric measurements are presented in Table~\ref{tab2}.
The apparent magnitude of Nereid is derived from the relative flux to the reference stars and their
total $V$ magnitude.
The brightness variation of a rotating body with such a small amplitude could be attributed to the
changes in the cross-section area and/or albedo inhomogeneity.
If Nereid is an ellipsoidal body with a constant albedo over the surface, it produces a
double-peaked sinusoidal lightcurve.
Indeed, \citet{Gr03} show that a simple sinusoid model gives a good fit to Nereid's lightcurve.
We assume that Nereid is covered by surface with a uniform albedo and its lightcurve is
doubly periodic.

For determination of the lightcurve, a variation of the observed magnitude ($m_{\rm obs}$) due to
the heliocentric distance ($r$ in AU) and geocentric distance ($\Delta$ in AU) must be corrected.
We use standardized magnitude ($m$)
derived from $m_{\rm obs} - 5 {\rm log}_{10}[r\Delta/900]$.
The rotational periodicity is analyzed by a Fourier-transform method called the Lomb-Scargle
periodogram \citep{Lo76,Sc82}.
The lightcurve model is represented as a first-order Fourier series formulation.
In addition, we approximated the opposition effect as linear increasing with approaching the
opposition.

The synthetic lightcurve with a period $P$ to be fitted is
\begin{equation}
m(t) = \frac{A_{\rm c}}{2} \cos \left( \omega (t - t_0) \right)
     + \frac{A_{\rm s}}{2} \sin \left( \omega (t - t_0) \right)
     + k \alpha + m_0,
\end{equation}
where $t$, $\omega$, $\alpha$, $k$, $m_0$ are the observed time, frequency ($\omega=2\pi/P$), solar
phase angle, slope of the phase curve, and basement magnitude, respectively.
The peak-to-peak amplitude $A$ is given by the coefficients $A_{\rm c}$ and $A_{\rm s}$ as
$A = \sqrt{A_{\rm c}^2 + A_{\rm s}^2}$.
The time epoch $t_0$ is evaluated by
\begin{equation}
\tan(2 \omega t_0) = \frac{ \sum^{N}_{i=1} \sin(\omega t_i)}
                          { \sum^{N}_{i=1} \cos(\omega t_i)},
\end{equation}
where $N$ is the number of data points.

\begin{figure}
  \begin{center}
    \FigureFile(80mm,80mm){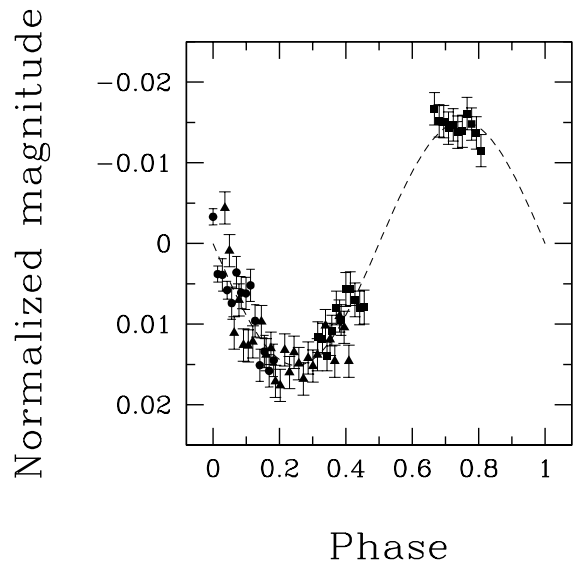}
  \end{center}
  \caption{
  Composite lightcurve of Nereid with a period of 5.75~hr.
  The vertical axis shows normalized magnitude corrected for the opposition effect.
  The circles, triangles and squares represent the data obtained on 2008 September 1, 2,
  and 29, respectively.
  The dashed curve is the best-fit model with a peak-to-peak amplitude of 0.031~mag.
  \label{fig1}}
\end{figure}

We obtained a best-fit model with
$P$ = 5.75 $\pm$ 0.05~hr,
$A$ = 0.031 $\pm$ 0.001~mag, and
$k$ = 0.138 $\pm$ 0.002~mag~deg$^{-1}$.
The mean residual of the fitting is 0.0026~mag, which is comparable to the photometric errors.
Figure \ref{fig1} shows the lightcurve combined with all the data points corrected for the
opposition effect and the best-fit model folded with the oscillation period $P$.
The rotation period (2$P$ = 11.50~hr $\pm$ 0.10~hr) and amplitude (0.031 $\pm$ 0.001~mag) well
agree with those presented by \citet{Gr03}, namely 11.52 $\pm$ 0.14~hr and 0.029 $\pm$ 0.003~mag.
This represents the constancy of Nereid's rotation state among August 2001, August 2002
\citep{Gr03}, and September 2008 (this study).
We found no evidence of the rotation variability over 7 years.
The rotation period is much shorter than the lower bound to induce the spin-orbit resonance or
chaotic rotation, about 2 weeks \citep{Do95}.
Nereid is observed to be and likely to remain in a constant rotation state.

Under the assumption of albedo homogeneity, the brightness fluctuation with 0.03-mag amplitude
constrains the body shape.
A triaxial ellipsoid with semi-axes $a \geq b \geq c$ rotating about the $c$ axis draws a
lightcurve with amplitude of
\begin{equation}
A = 2.5 \log \left( \frac{\bar{a}^2 \cos^2\theta + \bar{a}^2\bar{c}^2 \sin^2\theta}
                         {\bar{a}^2 \cos^2\theta + \bar{c}^2 \sin^2\theta} \right)^{1/2},
\end{equation}
where $\bar{a}=a/b$, $\bar{c}=c/b$, and $\theta$ is an aspect angle between the line of sight and
the spin axis \citep{LL03}.
Given a simple prolate spheroid ($i.e.$ $\bar{c}=1$), $\bar{a}$ $<$ 2.0 in $\theta > 15\arcdeg$.

For a prolate body, the period of precession derived from tidal torque is represented by
\begin{equation}
P_{\rm pre} \leq \frac{4(1-e^2)^{3/2}}{3(\bar{a}-1)} \frac{P^2_{\rm orb}}{P_{\rm rot}},
\end{equation}
where $P_{\rm pre}$, $P_{\rm orb}$, and $P_{\rm rot}$ are periods of the precession, orbit, and
rotation, respectively \citep{Do95,Sc08}.
It is already known that Nereid has $e$ = 0.75, $P_{\rm orb}$ = 360.13 days, and $P_{\rm rot}$ =
0.48 days (11.5~hr).
\citet{Sc08} suggest that Nereid's precessional period is $\sim$8 or $\sim$16 years.
Assuming $P_{\rm pre}$ = 16 years, the body must have $\bar{a}$ $\geq$ 4.3.
This elongated shape is hardly acceptable to reproduce a small lightcurve amplitude,
supporting the assertion that Nereid has a non-chaotic rotation with no precession.

It is widely known that atmosphereless bodies exhibit an exponential increase in reflectance at
tiny solar phase angle, called an opposition surge.
\citet{Sc08} determined the phase curves of Nereid using multi-year-combined photometric data on
246 nights from 1998 to 2006 (see also \cite{SS00,ST01}).
They have a sharp brightening exceeding 0.3~mag~deg$^{-1}$ down from $\alpha$ $\sim$
0.5$\arcdeg$, represented by the best-fit model for surface scattering \citep{Ha02} as
\begin{equation}
m = 19.655 - 2.5 {\rm log}_{10}\left\{ 1+0.54[1+(1-e^{-z})/z]/[2(1+z)^2]\right\},
\label{eq_pc}
\end{equation}
\begin{equation}
z=\tan[\alpha/2]/0.0134.
\end{equation}
On the other hand, \citet{Gr03} reported the surge slope of 0.14 $\pm$ 0.08 mag~deg$^{-1}$ at
$\alpha$ $\sim$ 0.4$\arcdeg$, which disagrees with the measurements by \citet{Sc08}.

\begin{figure}
  \begin{center}
    \FigureFile(120mm,120mm){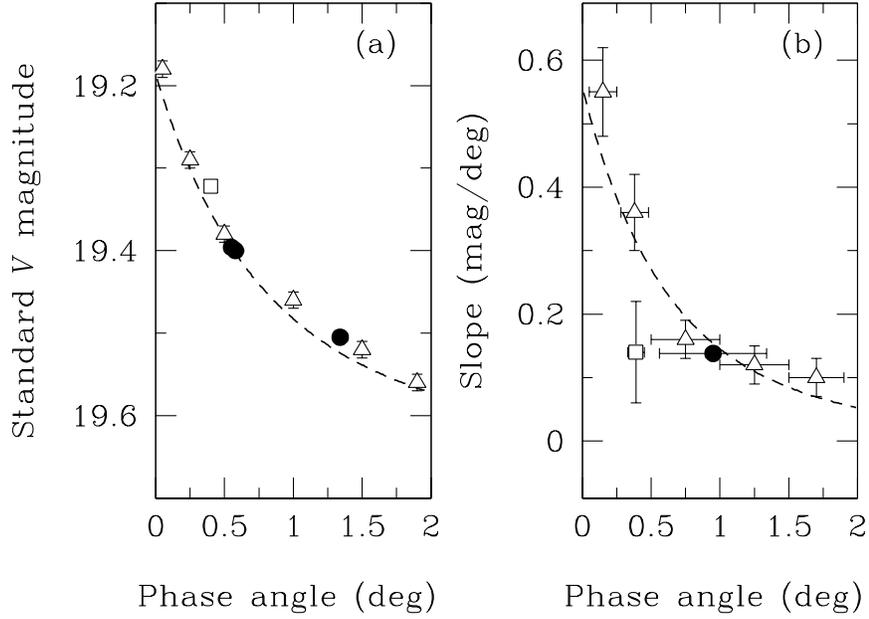}
  \end{center}
\caption{
 (a) $V$-band basement magnitude of Nereid's lightcurves standardized by heliocentric/geocentric
 distances ($m_{\rm obs} - 5 {\rm log}_{10}[r\Delta/900]$ ; see text) as a function of
 solar phase angle.
 The data are derived from this work (filled circles), \citet{Sc08} (open triangles), and \citet{Gr03}
 (open squares).
 The dashed curve is the best-fit model of the opposition surge given in \citet{Sc08}.
 (b) Slopes of the measured phase curves of Nereid in mag deg$^{-1}$.
 The symbols and dashed curve are the same as (a).
 The horizontal bar on each point shows the range of phase angle used in estimation of the slope.
 \label{fig2}}
\end{figure}

Figure~\ref{fig2} shows the combined phase curves and their slopes of Nereid
presented by this work and those of previous studies.
Both the magnitude and surge slope of our data are consistent with the model curve of
Eq.(\ref{eq_pc}).
In contrast, the surge given by \citet{Gr03} is too shallow.
The discrepancy is most likely due to the small range of phase angles spanned by their
observations (0.35$^{\circ}$--0.46$^{\circ}$).
There is no doubt that Nereid has a steep opposition surge, indicating that the coherent
backscattering mechanism is dominant on Nereid's surface \citep{Sc08}.

\section{DISCUSSION}

From the large, eccentric, and inclined orbit of Nereid, it is natural to consider its origin as
a captured body from outside the Neptune system.
\citet{Ne07} investigated the capture of irregular satellites in the planetesimal disk by
three-body gravitational reactions during encounters between the giant planets.
This model well reproduces the orbits of Nereid and most Neptunian irregular satellites.
Also, the long-term dynamical stability of Nereid's orbit has been confirmed \citep{Ho04}.
It is perfectly possible that Nereid has been an irregular satellite since it started orbiting
around Neptune.

\citet{Su11} found that long-lived temporary capture of planetesimals in the prograde direction
occurs at a limited range of initial orbital elements.
The capture rate has a peak at $\tilde{e} \simeq 3$ and $E \simeq 0$.
Here, $\tilde{e}$ is the initial orbital eccentricity scaled by $h = R_{\rm H}/a$, where
$R_{\rm H}$ is the Hill radius and $a$ is the semi-major axis of the planet;
$E$ is the energy integral described as
$E = (\tilde{e}^2 + \tilde{i}^2)/2 - 3\tilde{b}^2/8 + 9/2$,
where $\tilde{i}$ is the initial orbital inclination scaled by $h$ and $\tilde{b}$ is the difference
in the initial semi-major axis between the planet and a planetesimal scaled by $R_{\rm H}$
\citep{Na89}.
For Neptune, these conditions correspond to $e \simeq 0.08$ and $b \simeq$~3.8--4.6~AU in
$\tilde{i}$ ranging from 0 to $\tilde{e}$ ($e$ and $b$ are non-scaled $\tilde{e}$ and $\tilde{b}$,
respectively).
The source of Neptunian irregular satellites with prograde orbits is likely to be a planetesimal
population in the region between 25~AU and 35~AU if they were captured at the current location of
Neptune.

Although most planetesimals were lost from such a region, a portion of them possibly survives as
TNOs and Centaurs.
After dissipation of the disk gas, the giant planets underwent orbital migration due to angular
momentum exchange with residual planetesimals \citep{HM99,Ts05}.
This led to extensive outward transport of planetesimals from 25--35~AU to the Kuiper belt
\citep{Go03,LM03}.
\citet{LM03} indicate that small bodies in the entire Kuiper belt were formed within $\sim$35~AU
and were transported outward as a result of Neptune's migration.
Combining those studies, prograde irregular satellites of Neptune may have the same origin as TNOs
and Centaurs.
If Nereid derives from a captured planetesimal, this implication applies to it as well.

\begin{figure}
  \begin{center}
    \FigureFile(80mm,80mm){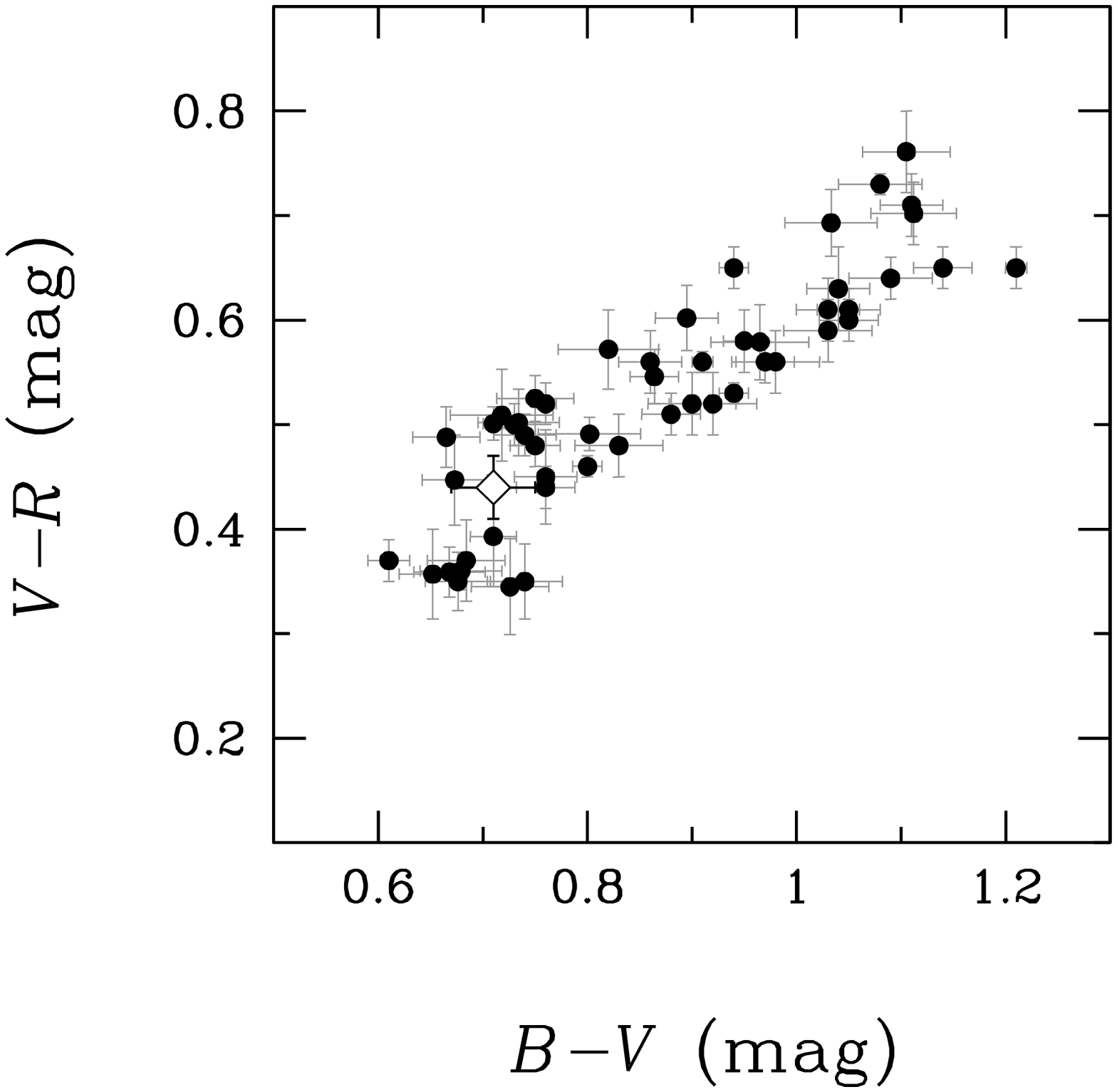}
  \end{center}
  \caption{
  Color-color diagram of $B-V$ vs. $V-R$ for Nereid (open diamond; \cite{SS00}) and TNOs/Centaurs
  (filled circles; \cite{HD02}).
  \label{fig3}}
\end{figure}

\begin{figure}
  \begin{center}
    \FigureFile(80mm,80mm){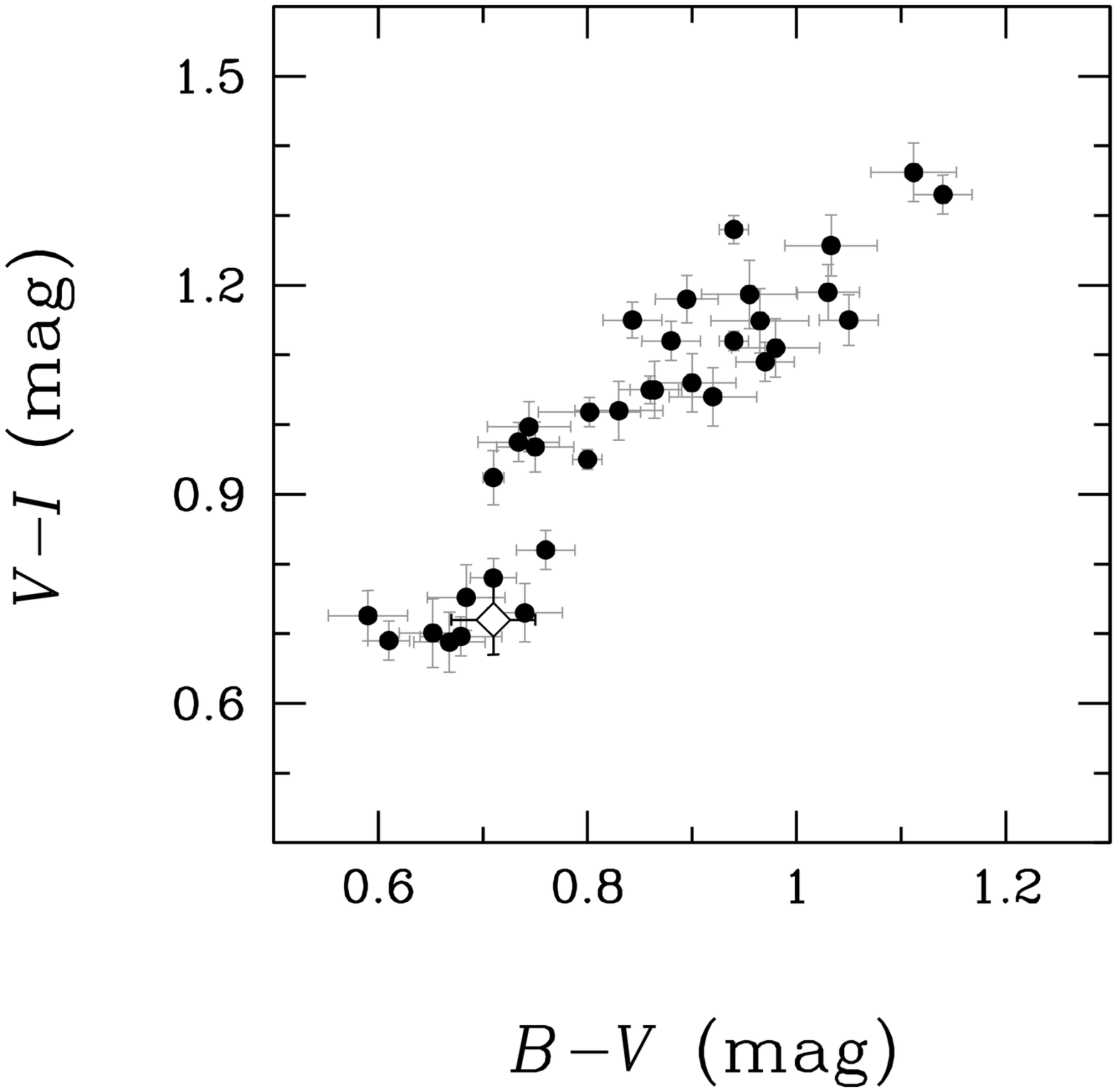}
  \end{center}
  \caption{
  Color-color diagram of $B-V$ vs. $V-I$ for Nereid (open diamond; \cite{SS00}) and TNOs/Centaurs
  (filled circles; \cite{HD02}).
  \label{fig4}}
\end{figure}

It is worth comparing Nereid with TNOs and Centaurs in term of some physical properties to examine
this hypothesis.
Nereid has optical colors of $B-V$ = 0.71 $\pm$ 0.04~mag, $V-R$ = 0.44 $\pm$ 0.03~mag, and
$V-I$ = 0.72 $\pm$ 0.05~mag \citep{SS00}.
The colors of 351 TNOs/Centaurs are provided by the MBOSS color
database\footnote{Minor Bodies in the Outer Solar System (http://www.eso.org/\~{}ohainaut/MBOSS).}
\citep{HD02}.
We use only those with uncertainties less than 0.05~mag for a comparison with Nereid.
Figure \ref{fig3} and \ref{fig4} show the color-color diagrams, $B-V$ vs. $V-R$ and $B-V$ vs.
$V-I$, respectively.
One can see that Nereid is indistinguishable from TNOs/Centaurs in the color-color planes,
supporting a common origin between them.
It is notable that Nereid's colors are similar to those of neutrally colored bodies rather than red
bodies.
\citet{Br11} suggest that TNOs formed in the inner part of the primordial disk retain H$_{2}$O,
CO$_{2}$ and poor hydrocarbon species on the surface which lead to a neutral color due to UV and
high-energy particle irradiation.
Nereid may have been originally formed at the region inside of $\sim$20~AU and transported into
near Neptune's orbit.

We also focus on the distribution of rotation period as another indicator of similarity between
Nereid and TNOs/Centaurs.
The rotation properties of small bodies, excluding tidally locked satellites, are dominantly
changed by mutual collisions (e.g. \cite{Ha79}).
The collisional evolution of rotation period depends on the size distribution, impact velocity,
and bulk density, which are unique among small-body populations.
Therefore, small-body populations with the same source are expected to have a common rotation
distribution.

\begin{figure}
  \begin{center}
    \FigureFile(80mm,80mm){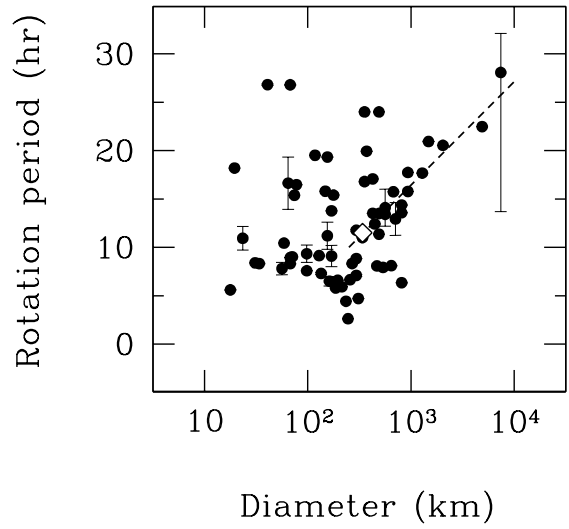}
  \end{center}
  \caption{
  The size-rotation distribution of TNOs and Centaurs presented in \citet{Du09}.
  The dashed line shows a linear regression line to the plot of $\log D$ ($D$ is diameter) vs.
  rotation periods of those bodies with $D$ $>$ 250~km except for Pluto, Charon, and Haumea.
  An open diamond represents Nereid.
  \label{fig5}}
\end{figure}

Interestingly, TNOs/Centaurs seem to have a characteristic pattern in the size-rotation
distribution.
Larger bodies rotate more slowly if they are greater than a few hundred kilometers in diameter
\citep{Sh08}.
We evaluate the probability of this trend using a large lightcurve database of 102 TNOs/Centaurs
compiled by \citet{Du09}.
Figure \ref{fig5} shows the distribution of their rotation periods against diameter derived by
an assumed albedo of 0.1.
There appears to be a monotonic increasing trend beyond 200--300~km.
We calculated the Spearman's rank correlation coefficient \citep{Sp04,Za72} of the sample including
33 objects with $D$ $\geq$ 250~km except for the tidally locked Pluto-Charon binary and Haumea, a
rapidly rotating dwarf planet.
The correlation coefficient is $r_{\rm s}$ = 0.53, meaning that large TNOs/Centaurs have a positive
correlation between rotation and size at the significance level greater than 99.5~\%.
This trend is explained by the random accumulation effect of angular momentum through
collisions \citep{Ha79,DB84}.

The rotation period of Nereid is plotted on Figure~\ref{fig5}.
We found that Nereid perfectly agrees with the size-rotation correlation of TNOs/Centaurs.
This agreement supports the idea that Nereid and TNO/Centaur populations share the same origin.
The capture scenario needs no additional spin-change events to explain the present rapid rotation,
unlike the other hypothesis based on the outward ejection from nearby Neptune.
It is more likely that Nereid is derived from a captured body rather than a primordial satellite of
Neptune.

\section{CONCLUSIONS}

We present highly accurate lightcurves of Nereid on two consecutive nights and one night after one
month later in September 2008 at phase angles of 0.5$\arcdeg$--1.3$\arcdeg$.
The periodic analysis shows that the rotation period of 11.5 $\pm$ 0.1~hr, peak-to-peak amplitude
in brightness variation of 0.031 $\pm$ 0.001~mag.
They completely agree with the measurements in \citet{Gr03}.
On the other hand, the steepness of magnitude increases due to the opposition surge is
0.138 $\pm$ 0.002~mag~deg$^{-1}$ which is consistent with the phase curve model presented by
\citet{Sc08}.
The conclusions of this work can be summarized as follows:

\begin{enumerate}

 \item The period and amplitude of Nereid's rotation are constant in 2001--2002 and 2008.
 We found no evidence of variability of the rotation state.
 The rapid rotation with a period much shorter than 2 weeks also rejects the chaotic rotation
 state suggested by \citet{Do95}.
 These facts indicate that Nereid has a stable rotation without a tidal despinning effect.
 
 \item Long-lived prograde captures of planetesimals by planets occur in a narrow range of
 the initial eccentricity and orbital energy \citep{Su11}.
 If Nereid was a body captured from a heliocentric orbit, the source region is likely to be
 from $\sim$25~AU to $\sim$35~AU where it is suggested that TNOs have been formed
 \citep{Go03,LM03}.
 
 \item The optical colors of Nereid are consistent with those of TNOs/Centaurs.
 Also, Nereid accords closely with the size-rotation distribution of TNOs/Centaurs.
 The available observations support the hypothesis that Nereid derived from a captured body and
 shares a common origin with the TNO/Centaur population.

\end{enumerate}

\bigskip

We thank David Jewitt for helpful comments.
This study is based in part on data collected at Subaru Telescope.
T. Terai was supported by the Grant-in-Aid from Japan Society for the Promotion of Science
(20-4879).


\clearpage


\begin{table}
  \caption{Time-series photometric observations for Nereid.}\label{tab1}
  \begin{center}
    \begin{tabular}{cccccccc}
      \hline
      Date        & RA2000   & Dec2000  & $r$\footnotemark[$*$] &
      $\Delta$\footnotemark[$\dagger$] & $\alpha$\footnotemark[$\ddagger$] &
      $T_{\rm obs}$\footnotemark[$\S$] &
      Number of\\
      (UT)        &          &          & (AU) & (AU) & (deg) & (hour) & usable images \\
      \hline
      2008 Sep 01 & 21$^{\rm h}$39$\fm$4  & -14$\arcdeg$22$\farcm$3 & 30.02 & 29.05 & 0.55 & 1.05 & 14 \\
      2008 Sep 02 & 21$^{\rm h}$39$\fm$3  & -14$\arcdeg$22$\farcm$8 & 30.02 & 29.06 & 0.58 & 2.15 & 26 \\
      2008 Sep 29 & 21$^{\rm h}$37$\fm$0  & -14$\arcdeg$34$\farcm$6 & 30.01 & 29.29 & 1.34 & 1.61 & 22 \\
      \hline
      \multicolumn{8}{@{}l@{}}{\hbox to 0pt{\parbox{85mm}{\footnotesize
        \par\noindent
        \footnotemark[$*$] Heliocentric distance.
        \par\noindent
        \footnotemark[$\dagger$] Geocentric distance.
        \par\noindent
        \footnotemark[$\ddagger$] Phase angle.
        \par\noindent
        \footnotemark[$\S$] Time length of the observation.
      }\hss}}
    \end{tabular}
  \end{center}
\end{table}

\begin{longtable}{ccc}
  \caption{Photometric measurements of Nereid.}\label{tab2}
  \hline
  MJD\footnotemark[$*$] & Airmass & $m_{V}$\footnotemark[$\dagger$] (mag) \\
  \hline
  \endhead
  \hline
  \endfoot
  \hline
  \multicolumn{3}{l}{\hbox to 0pt{\parbox{180mm}{\footnotesize
    \par\noindent
    \footnotemark[$*$] Modified Julian date.
    \par\noindent
    \footnotemark[$\dagger$] Apparent $V$-band magnitude.
  }}}
  \endlastfoot
  54710.2722 & 1.64 & 19.367 $\pm$ 0.001 \\
  54710.2755 & 1.61 & 19.374 $\pm$ 0.001 \\
  54710.2789 & 1.58 & 19.374 $\pm$ 0.002 \\
  54710.2823 & 1.55 & 19.376 $\pm$ 0.001 \\
  54710.2857 & 1.52 & 19.378 $\pm$ 0.002 \\
  54710.2891 & 1.50 & 19.374 $\pm$ 0.002 \\
  54710.2925 & 1.47 & 19.377 $\pm$ 0.002 \\
  54710.2959 & 1.45 & 19.377 $\pm$ 0.002 \\
  54710.2992 & 1.43 & 19.376 $\pm$ 0.002 \\
  54710.3026 & 1.41 & 19.380 $\pm$ 0.002 \\
  54710.3059 & 1.39 & 19.386 $\pm$ 0.002 \\
  54710.3093 & 1.37 & 19.384 $\pm$ 0.002 \\
  54710.3127 & 1.36 & 19.386 $\pm$ 0.002 \\
  54710.3160 & 1.34 & 19.385 $\pm$ 0.002 \\
  54711.2394 & 2.06 & 19.370 $\pm$ 0.002 \\
  54711.2427 & 2.00 & 19.376 $\pm$ 0.002 \\
  54711.2461 & 1.94 & 19.386 $\pm$ 0.002 \\
  54711.2495 & 1.89 & 19.382 $\pm$ 0.002 \\
  54711.2529 & 1.84 & 19.388 $\pm$ 0.002 \\
  54711.2563 & 1.79 & 19.388 $\pm$ 0.002 \\
  54711.2596 & 1.75 & 19.387 $\pm$ 0.002 \\
  54711.2658 & 1.68 & 19.385 $\pm$ 0.002 \\
  54711.2692 & 1.64 & 19.389 $\pm$ 0.002 \\
  54711.2726 & 1.61 & 19.388 $\pm$ 0.002 \\
  54711.2759 & 1.58 & 19.392 $\pm$ 0.002 \\
  54711.2793 & 1.55 & 19.393 $\pm$ 0.002 \\
  54711.2826 & 1.52 & 19.388 $\pm$ 0.002 \\
  54711.2860 & 1.50 & 19.391 $\pm$ 0.002 \\
  54711.2893 & 1.48 & 19.389 $\pm$ 0.002 \\
  54711.2927 & 1.45 & 19.390 $\pm$ 0.002 \\
  54711.2961 & 1.43 & 19.392 $\pm$ 0.002 \\
  54711.2995 & 1.41 & 19.389 $\pm$ 0.002 \\
  54711.3029 & 1.39 & 19.390 $\pm$ 0.002 \\
  54711.3062 & 1.38 & 19.389 $\pm$ 0.002 \\
  54711.3120 & 1.35 & 19.385 $\pm$ 0.002 \\
  54711.3153 & 1.33 & 19.387 $\pm$ 0.002 \\
  54711.3187 & 1.32 & 19.390 $\pm$ 0.002 \\
  54711.3221 & 1.31 & 19.384 $\pm$ 0.002 \\
  54711.3255 & 1.30 & 19.386 $\pm$ 0.002 \\
  54711.3289 & 1.29 & 19.390 $\pm$ 0.002 \\
  54738.2333 & 1.37 & 19.480 $\pm$ 0.002 \\
  54738.2367 & 1.35 & 19.481 $\pm$ 0.002 \\
  54738.2400 & 1.34 & 19.482 $\pm$ 0.002 \\
  54738.2434 & 1.32 & 19.482 $\pm$ 0.002 \\
  54738.2468 & 1.31 & 19.482 $\pm$ 0.002 \\
  54738.2501 & 1.30 & 19.483 $\pm$ 0.002 \\
  54738.2535 & 1.29 & 19.483 $\pm$ 0.002 \\
  54738.2568 & 1.28 & 19.481 $\pm$ 0.002 \\
  54738.2601 & 1.27 & 19.482 $\pm$ 0.002 \\
  54738.2635 & 1.26 & 19.483 $\pm$ 0.002 \\
  54738.2669 & 1.25 & 19.485 $\pm$ 0.002 \\
  54738.3888 & 1.42 & 19.509 $\pm$ 0.002 \\
  54738.3922 & 1.44 & 19.509 $\pm$ 0.002 \\
  54738.3955 & 1.46 & 19.511 $\pm$ 0.002 \\
  54738.3989 & 1.48 & 19.508 $\pm$ 0.002 \\
  54738.4022 & 1.50 & 19.505 $\pm$ 0.002 \\
  54738.4056 & 1.53 & 19.507 $\pm$ 0.002 \\
  54738.4090 & 1.55 & 19.503 $\pm$ 0.002 \\
  54738.4123 & 1.58 & 19.503 $\pm$ 0.002 \\
  54738.4156 & 1.61 & 19.504 $\pm$ 0.002 \\
  54738.4190 & 1.65 & 19.505 $\pm$ 0.002 \\
  54738.4223 & 1.68 & 19.505 $\pm$ 0.002 \\
\end{longtable}

\end{document}